\documentclass[conference]{IEEEtran}
\IEEEoverridecommandlockouts
\usepackage{authblk}

\usepackage{cite}
\usepackage{amsmath,amssymb,amsfonts}
\usepackage{algorithmic}
\usepackage{graphicx}
\usepackage{url}
\usepackage{textcomp}
\usepackage[table,x11names,dvipsnames,table]{xcolor}
\usepackage{graphicx,balance,multirow,xspace,subcaption}

\usepackage{pifont}
\newcommand{\cmark}{\ding{51}}%
\newcommand{\xmark}{\ding{55}}%
\makeatletter
\newcommand{\linebreakand}{%
  \end{@IEEEauthorhalign}
  \hfill\mbox{}\par
  \mbox{}\hfill\begin{@IEEEauthorhalign}
}
\makeatother

\def\BibTeX{{\rm B\kern-.05em{\sc i\kern-.025em b}\kern-.08em
    T\kern-.1667em\lower.7ex\hbox{E}\kern-.125emX}}

\newcommand{\numexp}{1,000\xspace}
\newcommand{\blocker}{\emph{IoTrimmer}\xspace}

\begin{document}

\title{Towards Automatic Identification and Blocking of\\ Non-Critical IoT Traffic Destinations}

\author[1]{Anna Maria Mandalari}
\author[1]{Roman Kolcun}
\author[1]{Hamed Haddadi}
\author[2]{Daniel J. Dubois}
\author[2]{David Choffnes}

\affil[1]{Imperial College London}
\affil[2]{Northeastern University}

%\author{
%  \IEEEauthorblockN{Anna Maria Mandalari, Roman Kolcun, Hamed Haddadi}
%  \IEEEauthorblockA{Imperial College London}
%%  \and
%%  \IEEEauthorblockN{Roman Kolcun}
%%  \IEEEauthorblockA{Imperial College London}
%%  \and
%%  \IEEEauthorblockN{Hamed Haddadi}
%%  \IEEEauthorblockA{Imperial College London}
%  \linebreakand % <------------- \and with a line-break
%  \IEEEauthorblockN{Daniel J. Dubois, David Choffness}
%  \IEEEauthorblockA{Northeastern University}
%%  \and
%%  \IEEEauthorblockN{David Choffnes}
%%  \IEEEauthorblockA{Northeastern University}
%}

\maketitle

\begin{abstract}

The consumer Internet of Things (IoT) space has experienced a significant rise in popularity in the recent years.
From smart speakers, to baby monitors, and smart kettles and TVs, these devices are increasingly found in households around the world while users may be unaware of the risks associated with owning these devices.
Previous work showed that these devices can threaten individuals' privacy and security by exposing information online to a large number of service providers and third party analytics services.
Our analysis shows that many of these Internet connections (and the information they expose) are neither critical, nor even essential to the operation of these devices. However, \emph{automatically} separating out critical from non-critical network traffic for an IoT device is nontrivial, and requires expert analysis based on manual experimentation in a controlled setting.

In this paper, we investigate whether it is possible to automatically classify network traffic destinations as either critical (essential for devices to function properly) or not, hence allowing the home gateway to act as a selective firewall to block undesired, non-critical destinations. 
Our initial results demonstrate that some IoT devices contact destinations that are not critical to their operation, and there is no impact on device functionality if these destinations are blocked.
We take the first steps towards designing and evaluating \blocker, a framework for automated testing and analysis of various destinations contacted by devices, and selectively blocking the ones that do not impact device functionality.

\end{abstract}

\if 0
The Internet of Things (IoT) has exploded in recent years, and it's not decreasing at any point in the near future.
Many IoT devices can be found in people's homes enabling popular applications such as TVs, smart speakers, and surveillance. 
Recent studies demonstrated that such IoT devices may expose information unexpectedly compromising users safety and privacy (e.g., motion secretly transmitted by a surveillance device), 
communicating to external destinations which are not critical or essential to their operations.
Users are often not aware of the operations of such devices at home, e.g., leakage of information, distributed attacks at the internal network.

In this paper, we design and implement a tool for tracking contacted destinations and limiting unnecessary communications.
The \blocker works as a network firewall that makes users aware of how the IoT devices operate in their home and also inform them about potential risks.
We demonstrate that the IoT devices contact at least one destination not used for the standard activity of the device. 
We discuss potential risks and future work to be done in this area.

\fi

% !TEX root = paper.tex
\section{Introduction}
\label{sec:introduction}

Consumer Internet of Things (IoT) devices are increasingly seen in our households, including smart TVs and speakers, surveillance cameras and doorbells, connected kitchen appliances, infrastructure monitoring devices, and wearables.
While these devices often come with interesting and beneficial services, they open the door to a variety of privacy and security risks.
Privacy risks, at the network and application layers, have been extensively covered in previous research~\cite{moniotr, 10.1145/3319535.3354198, 2019arXiv191103447V}.
IoT devices often contact a large number of destinations~\cite{moniotr} that can be classified as: \emph{First party}, manufacturer of the IoT device responsible for providing and supporting the device functionality; \emph{Support party}, any companies providing outsourced computing resources, such as CDN and cloud providers; and \emph{Third party}, any party that is not a first or support party, including advertising and analytics companies.

While a number of experimental and commercial \emph{IoT Security} solutions are available for blocking malicious or otherwise undesirable connections (see Section~\ref{sec:relatedwork} for more), these often rely on user configuration for each device, and often provide an \textit{all-or-nothing} connectivity option for traffic destinations without considering whether blocking traffic will break device functionality.  
%Moreover, IoT devices often have regular contact with a variety of services and establish network connections independently of their interaction with the users, making it extremely difficult to assess their benign versus malicious activities and destinations. 
There is a need for an approach that can block network traffic with little-to-no user configuration, and without breaking a device's primary functionality. Doing so can substantially reduce the privacy and security attack surfaces for IoT devices. A key requirement for this approach is to establish and maintain a list of network destinations that are essential for device functionality, and thus should not, in general, be blocked.

In this paper, we make a first attempt at providing an automated framework for detecting and isolating such non-critical communications from IoT devices.
We design and implement \blocker{}, a tool for rigorously testing the network connections made by each device against their functionality, and establishing whether limiting those specific connections adversely affects the device functionality.
In a way, \blocker{} works like a firewall or an ad blocker~\cite{198594} for IoT devices that can be implemented on a home gateway.
By using the device companion apps (on smartphones), \blocker{} is able to automatically control IoT devices and trigger their actions. 

Our key research contributions include:
\begin {itemize}
\item Designing and implementing automated experiments and self-validation for the interaction between IoT devices and their respective smartphone companion apps. 
\item Understanding DNS behavior for each device, i.e., collecting the set of destinations based on DNS requests.
\item Automatically detecting if such destinations are \emph{critical}, i.e., required for proper device functionality, and iteratively filtering the non-critical ones. 
\end {itemize}

As a proof-of-concept study for \blocker{}, we use three IoT devices: two security cameras and a smart bulb.
By running more than \numexp{} automated experiments, we find that some devices make more network connections than needed for them to function normally. 
There are some commonalities among the non-critical destinations contacted by different IoT devices:
some IoT devices contact non-critical destinations outside their region, and out of 9 destinations contacted by our three IoT devices, 4 are unnecessary.

While this short paper covers only a small number of devices at the time of writing, we are expanding our analysis to a much larger set. Based on our initial findings, we argue that our methodology has potential for: 
(i) Ability to evaluate \textit{portability} of the filter list from one device to other devices; 
(ii) Understanding \emph{partially}-necessary destinations (i.e. those necessary for one functionality but not for others); and 
(iii) evaluating blocking strategies and their impact on functionality.

\section{Goals and Assumptions}
\label{sec:assumptions}

 \begin{figure*}[t!]
  \centering
  \includegraphics[width=\linewidth]{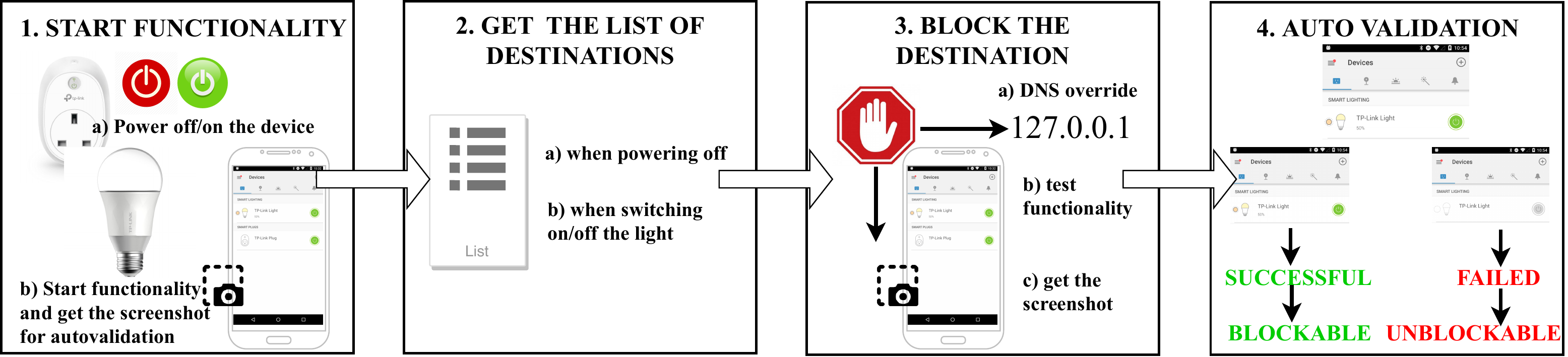}
  \caption{Overview of \blocker for the TP-Link smart bulb: (a) \textit{step 1}, power off/on the device and start functionality; \textit{step 2}, get the list of destinations; \textit{step 3}, block the destination using DNS override; \textit{step 4}, test the functionality and perform the auto validation.}
  \label{fig:blocker}
\end{figure*}

In this work we attempt to measure non-critical communication from popular consumer IoT devices to servers in the wild. 
In particular, we focus on characteristics of the destinations of their IP traffic, whether such communication is needed for the standard functionality of the device.
In this section, we define our key questions, challenges, and implications concerning destinations contacted by IoT devices. 

\subsection*{Goals}
\label{subsec:researchqs}
The goal of this work is to answer the following research questions.

\noindent \textbf{Do IoT devices communicate with destinations that are not critical to their functionality?} 
When a device initiates a communication with a destination over the Internet, it is by definition exposing some information. 
%Such information can be sent to first, support, and/or third parties. 
%Communication with support or third parties can be a privacy concern as personal information can be inferred. 
Since from our experience the vast majority of IoT traffic is encrypted, we do not know what information is actually
exposed. However, we can assume that if the device still works after blocking a destination, that destination
is unlikely to be essential to the functionality of the device (and thus the potential information exposure is not
needed). 
To answer this question, we track all the destinations that are not essential for the standard functionality of the device (i.e., considering a smart bulb, destinations that are not necessary for switching the light on/off).
From now on we refer to these destinations as \emph{blockable destinations}.

\noindent \textbf{Are there any patterns or trends on the blockable destinations among different devices?}
While there are a number of ways to classify destinations, in this short paper we focus only on whether any non-critical destinations are \emph{universal}---i.e., non-critical for different devices and categories. 
%Our work aims to detect destinations that are non-critical globally, for different devices and categories.

\subsection*{Assumptions}
\label{subsec:researchqs}
We consider as \textit{destination} the \textit{DNS destination}, the domain name requested by the device, according to its DNS queries.
We further distinguish destinations as follows: 
\begin{itemize}
\item \textit{Same domain, different ports} is considered a different destination.
\item \textit{Two different IPs}, resulting from resolving the same DNS destination, and contacted on the same port, are considered the same destination.
\item We use IP addresses in place of DNS name if there is no DNS name (i.e. hard-coded IP addresses).
\end{itemize}

%!TEX root = paper.tex

\section{\blocker Architecture}\label{sec:method}

We briefly introduce the mechanisms used by the \blocker to validate the automated experiments. Then, we describe our approach for understanding DNS behavior for consumer IoT devices and demonstrate the \blocker{} operations.

\subsection*{Testbed Setup and Devices}

To test our \blocker{}, we have built a testbed that currently comprises 122 different IoT devices in two labs, one in the US and one in the UK. 
We selected these devices to provide diversity within and between different categories: surveillance, video, audio, hub, appliance and home automation devices.
In addition to the devices, a Linux server running Ubuntu 18.04 with two Wi-Fi cards for 2.4\,GHz and 5\,GHz connections, plus two 1\,Gbps Ethernet connections for LAN and Internet connectivity are part of the setup. 
The server sits outside of any firewall and has a public IPv4 address. However, to match a regular home network environment, all IoT devices are behind a NAT setup and cannot be accessed directly from the Internet.

The monitoring software automatically detects connection of a new device to the network, assigns it a local IP address, and starts capturing packets using \emph{tcpdump}. 
Each device's traffic is filtered by MAC address into separate files.
The IoT devices can usually be controlled via a \emph{companion device} such as a smartphone application, an Alexa voice assistant or a Google Home. 
Our testbed allows us to perform automated experiments on the IoT devices using these companion devices.
In this case, the monitoring software captures the network traffic of both IoT and companion devices into separate PCAP files.
The testbed allows us to capture several network traces for each device, get destinations and perform self-validating automated experiments under different conditions.
%~\footnote{Our testbed code is available via \url{https://github.com/NEU-SNS/intl-iot/tree/master/moniotr}}
Finally, the testbed allows us to block traffic at device level by overriding DNS answers for specific hostnames using \emph{bind}'s view and RPZ zones.
%Finally, the testbed allows to block traffic at device level by either overriding DNS answers for specific hostnames using \emph{bind}'s RPZ zones, or blocking IP addresses using \emph{iptables}'s DROP rules.

\subsection*{Self-validating Automated Experiments}
\label{sec:val-exp}
The IoT devices can usually be controlled by a smartphone app or other IoT devices.
An empowering factor of \blocker{} is automation. 
Our \blocker{} involves automated interactions. 
For IoT devices that require a companion app, we use Nexus 5X smartphones running Android 6.0.1. 
We rely on the Monkey Application for Android Studio~\cite{monkey} for automating the interaction between the user and the IoT device. 
We conduct power experiments and automated interaction experiments to analyze the destinations contacted by each device under various conditions.

\noindent \textbf{Power experiments.} In our previous study~\cite{moniotr}, we found that many IoT devices transmit significant amount of traffic every time they are powered off and on. 
Based on this, the power experiments consist of controlling the devices through several TP-Link smart plugs that supports local control using a console application.
Thanks to these smart plugs, we are able to control the power status of the IoT devices under test, and power them off and on through scripts.
We force the devices to be off for a certain amount of time, then we turn them on and capture the network traffic for two more minutes with no operation. 
We automatically repeat the operation multiple times for every IoT device.

\noindent \textbf{Interaction experiments.} To detect the unnecessary destinations for an IoT device, we conduct \emph{interaction} experiments. 
An IoT device can have different functionalities, for example switching on/off the light for a smart bulb is the main functionality, but the app can also provide an option for controlling the brightness. 
To avoid mixing traffic from previous experiments, we perform interaction experiments every two minutes. 
After two minutes, and just before the communication begins, we start capturing the traffic and analyze the destinations contacted during the interactions required to trigger the device functionality.

Our interaction experiments include: (i) LAN application functionality, when the phone used for the interaction is on the same network as the IoT device, in which case the IoT device may communicate directly with the phone; (ii) WAN application functionality, when the phone is on another network than the IoT device. In this case the device must use a cloud service for communicating with the phone. 

The interaction between the IoT device and the phone app may fail. 
As part of the interaction experiments, we created a tool for validating them using screenshots matching. 
Screenshots may change because of dynamic data (i.e., time and temperature, appearance of pop up in the app, etc.), hence we only consider the part of the screen where the functionality is visible and crop that area. 
Figure~\ref{fig:blocker} Step 4 shows the methodology for such operation.

We perform the auto-validation in 6 steps: 
\begin{enumerate}
\item Automatically capturing all the screenshots of a given functionality of the companion app and use those screenshot as \textit{baseline} for comparison.
\item Starting the functionality automatically by opening the app using Monkey and sending the proper sequence of presses to trigger such functionality (i.e., switching on the light).
\item Using adb shell for taking the screenshot after the device interactions have been completed.
\item Transferring the screenshot from the phone to the processing server.
\item On the processing server, comparing the current screenshot with the \textit{base} screenshot, using ImageMagick.\footnote{\url{https://imagemagick.org/}}
\item If the screenshots match, we flag the experiment as successful, otherwise we consider it failed.
\end{enumerate}
 
\subsection*{DNS Behavior}
We created some experiments for understanding the DNS behavior for the IoT device.
We perform various power experiments, switching off the devices for 2sec, 4min, 8min, 16min, 32min, 64min, 128min, 188min.
This test helps us to understand the minimum duration to switch off a device before getting the list of destinations, in order to get as many DNS queries as possible.
  
\subsection*{Detecting and Blocking Destinations}
\blocker{} is able to automatically detect the list of destinations contacted by the IoT device and block the ones that are not necessary for the standard functionality of the device. 
Depending on the outcome of the self-validation, a destination can be either:
\begin{itemize}
\item \textbf{blockable for all experiments for that device}: blocking the destination does not break any functionality of the device.
\item \textbf{blockable for some experiments}: blocking the destination breaks some of the functionalities of the device.
\item \textbf{unblockable}: blocking the destination breaks all the functionalities of the device.
\end{itemize}

\blocker{} consists of two main steps, we (i) first gather the \emph{list the destinations} contacted by the device when powering off and on and while performing the functionality and (ii) secondly we \emph{block} the destinations one by one by testing the functionality using the auto validation tool. 

The steps are described below:
\begin{enumerate}
\item Powering off the device for two minutes. We capture the traffic for two minutes and get destinations.
\item Powering off the device again for two minutes. We start the functionality (i.e. switching on the light) and capture the traffic until the functionality execution is complete.
\item Performing the auto validation as shown in Figure~\ref{fig:blocker} and get the list of destinations contacted during power and interaction experiments.
\item Blocking the destination one by one by first powering off the device for two minutes. We block the destination using DNS override to map the DNS name to localhost (\emph{127.0.0.1}), effectively sending all traffic for that site to the loopback address.
\item Starting the functionality (i.e. switching on the light) and wait the necessary time until the functionality execution is complete.
\item Performing again the auto validation.
\item Based on the results of the auto validation, we flag the destination as either \textit{blockable} or \textit{unblockable} for that specific functionality.
\end{enumerate}

\subsection*{Repeating Experiments}
We repeat the experiments at least 30 times per device, per functionality.
In order to check the correct functionality of the experiment, \blocker{} sends a notification to us automatically, if the auto validation fails for more than three times in a row. 

%!TEX root = paper.tex
\section{IoTrimmer}
\label{sec:iot_blocker}

In this section, we study the destinations contacted by the IoT devices and then present the results from \blocker{}.
As an example, we consider three IoT devices from our testbed: TP-Link smart bulb, Yi Camera, and Bosiwo camera.

%!TEX root = ../paper.tex

\begin{table*}[!htbp]
	\centering
	\rowcolors{3}{gray!10}{white}
	
		\begin{tabular}{lr|cc|cc}
			\multirow{2}{*}{\textbf{Device}} & \multirow{2}{*}{\textbf{Activity}} & \multirow{2}{*}{\textbf{\# of }} & \multirow{2}{*}{\textbf{\# of }} & \multirow{2}{*}{\textbf{Domains/hard-coded IPs}}  	\\
			& & \textbf{domains} & \textbf{hard-coded IPs} & &  \\ \hline
%Meross door opener & LAN/WAN open, power  & 2  & 0 & ntp.org, iot.meross.com  \\
%\hline
TP-Link smart bulb & LAN/WAN switch off/on, power  & 5  & 0 & tplinkra.com, tplinkcloud.com, ntp.org, amazonaws.com, nist.gov\\
\hline
Yi Camera & LAN/WAN watch, power  & 2  & 0 & log.us.xiaoyi.com (WAN only), api.us.xiaoyi.com    \\
\hline
Bosiwo Camera & LAN/WAN watch, power  & 2 & 1  & vimtag.com, amazonaws.com, 210.72.145.44   \\
\hline
\end{tabular}

\caption{List of contacted destinations per device.} 
\label{table:domain}
%\posttabspace
\end{table*}
%\subsection*{Destination Analysis}
%\label{sec:destinations}

%In this section, we report the destinations contacted during the interaction and the power experiments by the devices. 

\noindent \textbf{Domain Destination Analysis.}
Table~\ref{table:domain} shows the list of domains contacted by the devices during the interactions and power experiments.
For example, TP-Link smart bulb contacts five domains during the power and the action of switching on/off the light: two NTP servers for synchronize the time, the TP-Link domains and the Amazon Web Server for the functionality of switching on/off the light. 

\noindent \textbf{IPs Destination Analysis and Hard-Coded IPs.}
Some devices may have hard-coded IPs. 
The IP address contacted is assigned manually in the settings of the device and does not change.
The Bosiwo camera for example contacted a hard-coded IP address. 
We use WHOIS data for identifying the owner of the IP address as reported by the corresponding regional registry.
 
% \begin{figure}
%  \centering
%  \includegraphics[width=\linewidth]{fig/dns}
%  \caption{Number of unique DNS queries per device, per duration of the power experiment.}
%  \label{fig:dns}
%\end{figure}

\noindent \textbf{DNS Behavior.} We study the number of unique DNS queries per device, per duration of the power experiment. 
Results show that the bulb made 5 queries and the cameras made 2 queries each, regardless of the duration of the power experiment.
The test validates that it is safe for each experiment to wait no more than 2 minutes for getting the destinations. 

%!TEX root = ../paper.tex

\begin{table}[t]
	\centering
	\rowcolors{3}{gray!10}{white}
		\resizebox{1.0\columnwidth}{!}{	
		\begin{tabular}{lr|ccc}
			\multirow{2}{*}{\textbf{Device}} & \multirow{2}{*}{\textbf{Activity}} & \multirow{2}{*}{\textbf{Domains }} & \multirow{2}{*}{\textbf{Blockable }}   	\\
			& &  &    \\ \hline
%Meross door opener & LAN open  & ntp.org  & \cmark  \\
%\hline
%Meross door opener & LAN open  & iot.meross.com  & \xmark  \\
%\hline
%Meross door opener & WAN open  & ntp.org  & \cmark  \\
%\hline
%Meross door opener & WAN open  & iot.meross.com  & \xmark  \\
%\hline
TP-Link smart bulb & LAN switch off/on  & tplinkra.com  & \xmark  \\
\hline
TP-Link smart bulb & LAN switch off/on  & tplinkcloud.com  & \xmark  \\
\hline
TP-Link smart bulb & LAN switch off/on  & ntp.org  & \cmark  \\
\hline
TP-Link smart bulb & LAN switch off/on  & amazonaws.com  & \xmark  \\
\hline
TP-Link smart bulb & LAN switch off/on  & nist.gov  & \cmark  \\
\hline
TP-Link smart bulb & WAN switch off/on  & tplinkra.com  & \xmark  \\
\hline
TP-Link smart bulb & WAN switch off/on  & tplinkcloud.com  & \xmark  \\
\hline
TP-Link smart bulb & WAN switch off/on  & ntp.org  & \cmark  \\
\hline
TP-Link smart bulb & WAN switch off/on  & amazonaws.com  & \xmark  \\
\hline
TP-Link smart bulb & WAN switch off/on  & nist.gov  & \cmark  \\
\hline
Yi Camera & LAN watch  &   api.us.xiaoyi.com & \xmark    \\
\hline
Yi Camera & WAN watch  &  api.us.xiaoyi.com & \xmark    \\
\hline
Yi Camera & WAN watch  & log.us.xiaoyi.com & \cmark    \\
\hline
Bosiwo Camera & LAN watch  & vimtag.com & \xmark    \\
\hline
Bosiwo Camera & LAN watch  & amazonaws.com & \xmark    \\
\hline
Bosiwo Camera & LAN watch  & 210.72.145.44 & \cmark    \\
\hline
Bosiwo Camera & WAN watch  & vimtag.com & \xmark    \\
\hline
Bosiwo Camera & WAN watch  & amazonaws.com & \xmark    \\
\hline
Bosiwo Camera & WAN watch  & 210.72.145.44 & \cmark    \\
\hline
\end{tabular}
}
\caption{List of blockable/unblockable destinations per device.} 
\label{table:block}
%\posttabspace
\end{table}

\noindent \textbf{Blockable Destinations.}
Table~\ref{table:block} shows the destinations contacted by the device and whether it is blockable or not~(\cmark, \xmark). 
Among 9 destinations contacted, 4 are blockable (i.e., blocking them does not affect the device functionality for that given experiment).
At least one destination is blockable for each device. 

Blockable destinations are mostly support parties, needed for clock synchronization (ntp.org, nist.org) and logs (log.us.xiaoyi.com), except for the hard-coded IP contacted by the Bosiwo camera. 
It is not immediately clear why the camera continuously pings that IP address. 
Our analysis shows that the owner of the IP is \emph{Computer Network Information Center} in China.
 
%!TEX root = ../paper.tex

\begin{table}[ht!]
	\centering
	\rowcolors{3}{gray!10}{white}
		\resizebox{1.0\columnwidth}{!}{	
		\begin{tabular}{lr|ccc}
			\multirow{2}{*}{\textbf{Destination}} & \multirow{2}{*}{\textbf{\#of Devices}} & \multirow{2}{*}{\textbf{Protocol}} & \multirow{2}{*}{\textbf{Port}} & \multirow{2}{*}{\textbf{Amount of Traffic(\%)}}   	\\
			& &  & &   \\ \hline
ntp.org & 1  & NTP  & 123  & 1.53  \\
\hline
nist.org & 1  & NTP  & 123 & 1.47\%  \\
\hline
log.us.xiaoyi.com & 1  & TCP & 80 & 2.8\%  \\
\hline
210.72.145.44 & 1  & ICMP  & 123 & 0.12\%  \\
\hline
\end{tabular}
}
\caption{Traffic characterization for the blockable destinations. For each destination we list the number of devices contacting that destination, the protocol, the destination port and the average amount of traffic (in percentage) sent to that destination over the total traffic sent for the entire duration of the experiment.} 
\label{table:traffic}
%\posttabspace
\end{table}

Table~\ref{table:traffic} summarizes the characteristics of the traffic sent to the destinations that are blockable.
The traffic includes NTP, TCP over port 80 (HTTP traffic), and ICMP protocols. 

 %\subsection*{General/Partially Blockable Destinations and Changes over Time}
\noindent \textbf{Generalization Analysis.}
We now investigate similarities among blockable destinations (e.g., same port, domain, organization, 2nd-level domain, etc.). 
%We also investigate the blockable destinations that are never unblockable. 
This helps to generalize the blocking from one to other devices. 
Results show that support services such as AmazonAWS are never blockable, while servers for NTP traffic are always blockable for all the experiments and all the devices.
Some destinations may be blockable for one functionality and unblockable for others, we do not detect that behavior.

\noindent \textbf{Longitudinal Analysis.}
We also perform experiments for understanding whether a destination is blockable at given time and not blockable anymore after a certain amount of time.
We repeat the experiments in different times for a period of one month. 
We did not find any significant differences on blockable destinations or destinations contacted.

%!TEX root = paper.tex
\section{Related Work}
\label{sec:relatedwork}

The increasing awareness in privacy and security risks in the consumer IoT market has led to a number of tools to protect against abnormal IoT traffic.  For example, SPIN\footnote{\url{https://spin4home.nl/}} is an open source software tool which focuses on visualizing and blocking traffic to and from IoT devices. 
Tools and platforms such as ShieldIOT,\footnote{\url{https://shieldiot.io/}} Fing,\footnote{\url{https://www.fing.com/}} and Bitdefender\footnote{\url{https://www.bitdefender.com/iot/}} provide solutions to protect against known vulnerabilities, or to isolate devices from the network based on their MAC/IP address.
However, the majority of these solutions depends on existing knowledge of devices and their destinations, which is not easy to obtain and evaluate with an ever-expanding set of devices.

There are a number of existing tools for IoT traffic and privacy risk analysis. 
For examples, IoT Inspector~\cite{huang2019iot} collects smart home traffic in scale using ARP spoofing. 
In recent work~\cite{moniotr}, we studied information exposure from 81 consumer IoT devices from two testbeds in different countries. 
%We similarly use data collected from these two testbeds for studying destinations and testing \blocker{}. 

One of our main objectives is to analyze and block destinations specifically for a given device, and for specific device-functionality scenarios.
\blocker{} tests the device functionality and blocks the destinations which are blockable without impacting other functionalities. 
\blocker{} will also help to define general rules for blocking unnecessary destinations in smart home environments, in manner similar to ad blockers in browsers~\cite{4939739}.

%Many tool for collecting smart traffic and traffic in general have been also implemented.

%We now review work related to IoT  and general traffic blocking techniques.

%Existing techniques for collecting and analyzing smart home traffic at the edge has been implemented recently.

%Bitdefender~\cite{bitdefender} and Fing~\cite{fing} created a boxes for security solution for the smart, connected home. 

%These commercial products may be install at home on the ISP provided router. They monitor traffic form all the devices at home to the Internet,
%block websites, look for anomalies due to the security network issues and much more. 

%However,  we are not aware of any software that analyze and block destinations specifically for one device and one device-functionality. 

%The tool Netalyzr~\cite{netalyzr} is a web application and mobile app that collect and analyze network traffic from mobile and home devices. 

%!TEX root = paper.tex

\section{Discussion and Conclusion}
\label{sec:discussion}

Having motivated the need for identifying non-critical destinations in IoT environments, we created and tested \blocker{}, an automated framework that filters those non-critical connections, testing automatically the device functionality.

We presented results from three IoT devices in our lab, two cameras and a smart bulb. We performed more than \numexp{} automated experiments and validated them.
Most devices make more network connections than needed for them to function normally. 
Out of 9 destinations contacted by three of our tested IoT devices, 4 are blockable. 
Some devices, such as the Bosiwo camera, contact destinations located in countries outside of their testbed's privacy jurisdiction.

In this paper we demonstrated that limiting NTP traffic does not affect the device functionality over the timescale of our study; however time desynchronization may affect other functionalities (e.g., certificate validation) in some cases.
As future work, we plan to analyze the cascading effects of limiting that kind of traffic, and alternatives to always blocking (e.g., allowing clock synchronization at randomized intervals).
We also plan to detect whether each non-critical destination is used for other important purposes (e.g., updating the firmware of the device).
In addition to exploring non-critical traffic for more devices and over longer timescales, we will investigate the potential for blocking traffic between devices in the same LAN, and study differences arising from deploying the same devices in different geographical regions.
Finally, we plan to implement and evaluate \blocker{} on an edge platform such as the Databox~\cite{iotdatabox} as a home gateway device.

%\noindent \textbf{Do different regions have different blockable destinations?} 
%IoT devices may contact different destinations depending on the device's location (jurisdiction, location of network egress).  
%\blocker{} can identify highlight regional differences for blockable/unblockable destinations.
%The main question will be: are the destinations \emph{regionally} undesired (e.g., undesired for the UK, but necessary for the US)? 

\section*{Acknowledgments}

We acknowledge constructive comments and feedback from the reviewers of IEEE S\&P ConPro'20. This research was partially supported by the NSF CNS-1909020, the EPSRC Databox  (EP/N028260/1), and the EPSRC Defence Against Dark Artefacts (EP/R03351X/1) grants.

\bibliographystyle{IEEEtran}
\bibliography{IEEEabrv,paper}

\begin{thebibliography}{1}
\providecommand{\url}[1]{#1}
\csname url@samestyle\endcsname
\providecommand{\newblock}{\relax}
\providecommand{\bibinfo}[2]{#2}
\providecommand{\BIBentrySTDinterwordspacing}{\spaceskip=0pt\relax}
\providecommand{\BIBentryALTinterwordstretchfactor}{4}
\providecommand{\BIBentryALTinterwordspacing}{\spaceskip=\fontdimen2\font plus
\BIBentryALTinterwordstretchfactor\fontdimen3\font minus
  \fontdimen4\font\relax}
\providecommand{\BIBforeignlanguage}[2]{{%
\expandafter\ifx\csname l@#1\endcsname\relax
\typeout{** WARNING: IEEEtran.bst: No hyphenation pattern has been}%
\typeout{** loaded for the language `#1'. Using the pattern for}%
\typeout{** the default language instead.}%
\else
\language=\csname l@#1\endcsname
\fi
#2}}
\providecommand{\BIBdecl}{\relax}
\BIBdecl

\bibitem{moniotr}
J.~Ren, D.~J. Dubois, D.~Choffnes, A.~M. Mandalari, R.~Kolcun, and H.~Haddadi,
  ``Information exposure from consumer iot devices: A multidimensional,
  network-informed measurement approach,'' in \emph{Proceedings of the Internet
  Measurement Conference}, 2019.

\bibitem{10.1145/3319535.3354198}
H.~Mohajeri~Moghaddam, G.~Acar, B.~Burgess, A.~Mathur, D.~Y. Huang,
  N.~Feamster, E.~W. Felten, P.~Mittal, and A.~Narayanan, ``Watching you watch:
  The tracking ecosystem of over-the-top tv streaming devices,'' in
  \emph{Proceedings of the 2019 ACM SIGSAC Conference on Computer and
  Communications Security}, 2019.

\bibitem{2019arXiv191103447V}
J.~Varmarken, H.~Le, A.~Shuba, A.~Markopoulou, and Z.~Shafiq, ``The tv is smart
  and full of trackers: Measuring smart tv advertising and tracking,''
  \emph{Proceedings on Privacy Enhancing Technologies}, vol.~2, pp. 129--154,
  2020.

\bibitem{198594}
R.~Nithyanand, S.~Khattak, M.~Javed, N.~Vallina-Rodriguez, M.~Falahrastegar,
  J.~E. Powles, E.~D. Cristofaro, H.~Haddadi, and S.~J. Murdoch, ``Adblocking
  and counter blocking: A slice of the arms race,'' in \emph{6th {USENIX}
  Workshop on Free and Open Communications on the Internet ({FOCI} 16)},
  Austin, TX, Aug. 2016.

\bibitem{monkey}
P.~Patel, G.~Srinivasan, S.~Rahaman, and I.~Neamtiu, ``On the effectiveness of
  random testing for android: Or how i learned to stop worrying and love the
  monkey,'' in \emph{Proceedings of the 13th International Workshop on
  Automation of Software Test}, 2018.

\bibitem{huang2019iot}
D.~Y. Huang, N.~Apthorpe, G.~Acar, F.~Li, and N.~Feamster, ``Iot inspector:
  Crowdsourcing labeled network traffic from smart home devices at scale,''
  2019.

\bibitem{4939739}
{Ashish Kumar Singh} and V.~{Potdar}, ``Blocking online advertising - a state
  of the art,'' in \emph{2009 IEEE International Conference on Industrial
  Technology}, Feb 2009, pp. 1--10.

\bibitem{iotdatabox}
\BIBentryALTinterwordspacing
{Crabtree et al.}, ``{Building accountability into the Internet of Things: the
  IoT Databox model},'' \emph{Journal of Reliable Intelligent Environments},
  vol.~4, no.~1, pp. 39--55, 2018. [Online]. Available:
  \url{https://doi.org/10.1007/s40860-018-0054-5}
\BIBentrySTDinterwordspacing

\end{thebibliography}

\end{document}